# X-ray linear dichroic orientation tomography: reconstruction of nanoscale three-dimensional orientation fields

Andreas Apseros*,[1,2] Valerio Scagnoli*,[1,2] Manuel Guizar-Sicairos,[3,4] Laura J. Heyderman,[1,2] Johannes Ihli,[5] Claire Donnelly*,[6,7]

[1]Laboratory for Mesoscopic Systems, Department of Materials, ETH Zurich, 8093 Zurich, Switzerland
[2]PSI Center for Neutron and Muon Sciences, 5232 Villigen PSI, Switzerland
[3]PSI Center for Photon Science, 5232 Villigen PSI, Switzerland
[4]École Polytechnique Fédérale de Lausanne, 1015 Lausanne, Switzerland
[5]University of Oxford, OX2 6NN Oxford, United Kingdom
[6]Max Planck Institute for Chemical Physics of Solids, 01187 Dresden, Germany
[7]International Institute for Sustainability with Knotted Chiral Meta Matter (WPI-SKCM[2]), Hiroshima University, Hiroshima 739-8526, Japan

* Correspondence and requests for materials should be addressed to andreas.apseros@psi.ch, valerio.scagnoli@psi.ch, claire.donnelly@cpfs.mpg.de

## Abstract

Properties in crystalline and ordered materials tend to be anisotropic, with their orientation affecting the macroscopic behavior and functionality of materials. The ability to image the orientation of anisotropic material properties in three dimensions (3D) is fundamental for the understanding and functionality-driven development of novel materials. With the development of X-ray linear dichroic orientation tomography (XL-DOT), it is now possible to non-destructively map three-dimensional (3D) orientation fields in micrometer-sized samples. In this work, we present the iterative, gradient-based reconstruction algorithm behind XL-DOT that can be used to map orientations based on linear dichroism in 3D. As linear dichroism can be exhibited by a broad spectrum of materials, XL-DOT can be used to map, for example, crystal orientations as well as ferroic alignment, such as ferroelectric and antiferromagnetic order. We demonstrate the robustness of this technique for orientation fields that exhibit smoothly varying and granular configurations, and subsequently identify and discuss optimal geometries for experimental data acquisition and optimal conditions for the reconstruction. We anticipate that this technique will be instrumental in enabling a deeper understanding of the relationship between material structures and their functionality, quantifying, for example, the orientation of charge distributions and magnetic anisotropies at the nanoscale in a wide variety of systems – from functional to energy materials.

# 1. Introduction

Orientation fields and anisotropy on the nanoscale are present in many systems and determine the macroscopic structure, mechanical properties and functionality of materials[1-4]. For example, local crystallographic microstructure strongly influences mechanical properties and can determine key properties such as the coercive fields of permanent magnets[5] and ferroelectrics[6], as well as the stability and efficiency of batteries[7,8] and catalysts[9,10]. In addition, the local orientation of ensembles of molecules is key to determining the properties of biological materials such as bone[11,12]. Furthermore, orientation fields that form in ferroic materials such as the Néel vector in antiferromagnets[13,14], or the polarization vector in ferroelectrics, provide possibilities to store and process data in new and efficient ways.

Until now, the mapping of 3D orientation fields at the nanoscale has been challenging. Electron-based techniques offer high spatial resolution maps[15-17] — with spatial resolutions on the order of a few nanometers or below — but rely on sample-destructive serial sectioning methods to obtain 3D information from micrometer-sized samples[18,19]. Synchrotron X-ray diffraction techniques can also reach nanometer resolution in 3D, in particular when using Bragg coherent diffractive imaging[20-22], providing information on strain and charge distributions. However, with Bragg coherent diffractive imaging, the sample sizes tend to be similarly constrained to the nanoscale. The imaging of orientations in 3D in extended polycrystalline and biological samples has also been achieved with diffraction contrast tomography[23,24] and scanning SAXS tensor tomography[25], respectively, albeit with a spatial resolution on the order of a few hundred nanometers.

With the development of X-ray linear dichroic orientation tomography[26] (XL-DOT), the nanoscale mapping of orientation fields — ranging from crystallographic orientations to other systems exhibiting linear dichroism such as ordered biological materials[1,11,12], antiferromagnetic[13,14] and ferroelectric[6] systems — now becomes possible. The technique relies on three basic requirements: (i) the measurement of two-dimensional transmission projections of the orientation field with nanoscale spatial resolution, (ii) the measurement of sufficient data to probe and successfully recover all components of the orientation field and (iii) a dedicated reconstruction algorithm to recover the 3D configuration from the linear dichroic data.

In this work, we focus on the third requirement and present the gradient-based iterative tomographic reconstruction algorithm of XL-DOT, that allows for the robust reconstruction of 3D orientation fields from X-ray linear dichroic data. We test the validity of the algorithm against different types of orientation configurations by simulating XL-DOT for smoothly varying and granular model systems, and determine the optimal measurement conditions for the effective use of this technique, identifying certain conditions that might limit the accuracy of the reconstruction. Finally, we explore the impact of known experimental limitations such as reduced sample tilting capabilities, or other constrained experimental geometries, in order to characterize the impact that measurement conditions may have on the ability to reconstruct 3D orientation fields.

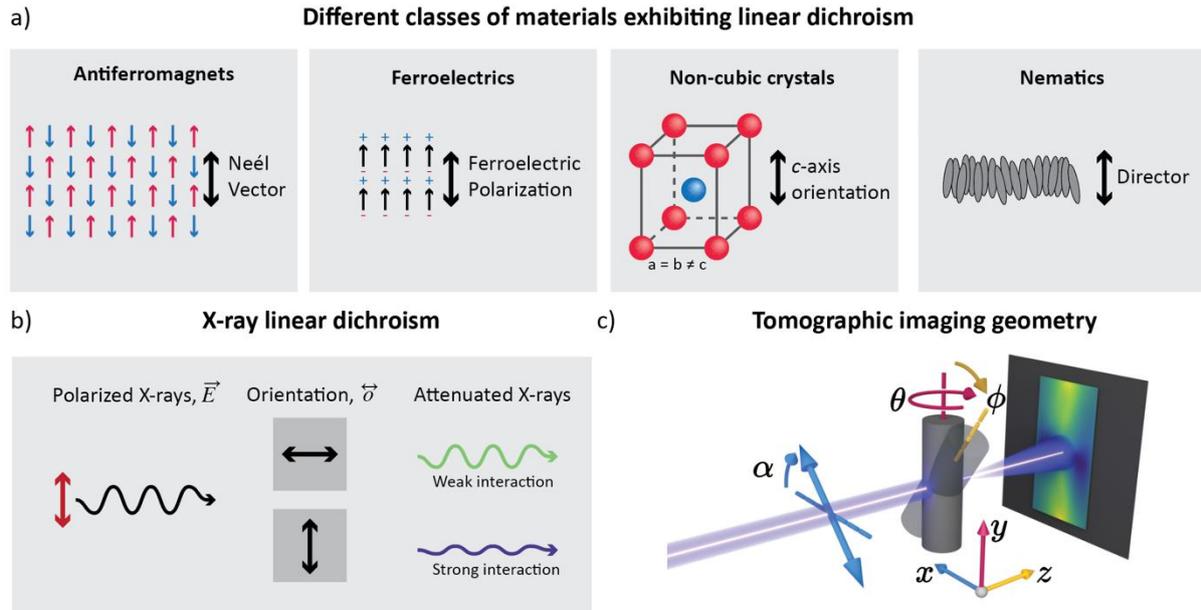

**Figure 1**: a) Illustration of different classes of materials exhibiting linear dichroism. These include antiferromagnets, ferroelectrics, non-cubic crystals and liquid crystals. The orientation of the probed anisotropy is represented using a double-headed arrow to emphasize the loss of directional information. b) The strength of the linear dichroic interaction between the linearly polarized X-rays and the material depends on the orientation of the polarization of the incident beam $\vec{E}$ relative to the orientation of the probed order parameter, $\vec{o}$. In the illustration, the X-rays are strongly absorbed and attenuated when $\vec{E}$ and $\vec{o}$ are parallel, and the X-rays are weakly absorbed when the two are perpendicular. c) Tomographic imaging comprising linear dichroic measurements is used to map the probed orientation in 3D. In tomography, the sample is tilted about the $z$ axis by an angle $\phi$ and is then rotated about the $y$ axis, varying the tomographic rotation angle $\theta$ from 0° to 180° while recording projections at regular angular intervals. The blue double-headed arrow represents the linear polarization, where $\alpha$ is the polarization angle relative to the $x$ axis. Multiple tilt angles $\phi$ tend to be necessary for vector and orientation tomography.

## 2. Tomographic implementation of X-ray linear dichroism

### 2.1 Overview of X-ray magnetic and natural linear dichroism

X-ray linear dichroism refers to the change in the strength of the interaction between the incident X-rays and the material when the orientation of the linear polarization of the X-rays is changed. X-ray linear dichroism is typically classified into magnetic linear dichroism (XMLD) and natural linear dichroism (XLD). XMLD arises in magnetic materials where the spins are aligned[27,28] and can be used to probe the orientation of the magnetic alignment axis. For instance, it can be used to determine the orientation of the Néel vector in antiferromagnetic materials as illustrated in the first panel in Fig. 1a. In contrast, XLD arises from the anisotropic charge distribution inside materials[28]. This anisotropic charge distribution can arise, for instance, as a result of the displacement of ions in non-centrosymmetric crystal lattices, which can give rise to ferroelectricity, or directional bonds that break the symmetry in an otherwise cubic crystal, illustrated in the two middle panels in Fig. 1a. Using X-ray resonant scattering and tuning the X-ray energy to specific electronic transitions in the material, it is possible to exploit linear dichroism to probe the charge anisotropy in the valence orbitals of the sample. The total scattering factors, $f$, for a linear dichroic interaction can generally be expressed as[1,28]

$$f = f_0 + f_{\text{lin}}\left(\vec{E} \cdot \vec{o}\right)^2, \tag{1}$$

with the prefactors $f_0$ and $f_{\text{lin}}$ denoting the conventional electronic and linear dichroism scattering factors, respectively. $\vec{E}$ is the electric field unit vector associated with the incident X-rays, in other words the polarization, and $\vec{o}$ is the local orientation of the order parameter in the material. The magnitudes of these scattering factors depend on the cross-sections of the associated resonant electronic transitions. We note that the $f_0$ term is non-dichroic because it is independent of the orientation of the incident polarization, whereas the magnitude of the second term varies with the dot product between the polarization vector and the orientation probed, $\vec{E} \cdot \vec{o}$. The squaring of this term takes into account the fact that it is not possible to distinguish the order parameter direction, which we emphasize here by labeling it with a double-headed arrow, $\vec{o}$. Consistent with the dot product present in Equation 1, XLD contrast is high when the polarization vector, $\vec{E}$, is collinear with the orientation, $\vec{o}$, and low when the two quantities are perpendicular, as illustrated in Fig. 1b.

Since its prediction and experimental confirmation, XLD has been extensively exploited to image magnetic and charge anisotropies in thin film systems using X-ray photoemission electron microscopy (X-PEEM)[29] or projections of extended systems using (Scanning) Transmission X-ray Microscopy (S)TXM[30]. Examples include the orientation of the polarization in ferroelectrics[6], the Néel vector in antiferromagnets[13,14], the director in liquid crystals[1,31] as well as the orientation of proteins[11] and biominerals[12] in organic samples. However, while circular dichroism has been exploited for the 3D mapping of magnetic vector fields[32,33], mapping of the orientation in 3D with linear dichroism — something that in many cases is not accessible with circular dichroism — has never been implemented. Here, we elaborate on the recent development of XL-DOT[26] for the 3D mapping of orientation fields, demonstrating its applicability to visualize ordering across various materials.

## 2.2 Theory of Tomographic Imaging

XL-DOT combines tomographic imaging with the information of the orientation of the magnetic or charge anisotropy probed using linear dichroism to provide a 3D map of the order parameter. To perform 3D vector orientation or tensor imaging, typically more data is required than for a standard tomographic measurement. For the case of magnetic vector tomography, with circularly polarized X-rays for example, successful reconstructions have been obtained using both multi-tilt tomography[32,33] and laminography geometries[34,35].

In this work, we determine the necessary and optimal geometries for acquiring data to perform XL-DOT and consider the impact of limited datasets. To maintain flexibility in terms of the experimental geometry, we develop our method for arbitrary geometries by describing the 3D imaging process in terms of rotation matrices[33].

Specifically, we can express the linear dichroic projections $P(x, y)$ of the 3D orientation field $\vec{o}$ as a function of the sample rotation and tilt, $\mathbf{R}$, and the incident polarization, $\vec{E}$:

$$P(x,y) = \int f_0(R_{ji}r_i) + f_{\text{lin}}\big(R_{ij}o_j(R_{ji}r_i)E_i\big)^2 \mathrm{d}z \,. \tag{2}$$

Here, the coordinate system is such that the $z$ axis is parallel to the propagation direction of the X-ray beam, the $x$ axis is horizontal, and the $y$ axis is vertical, as shown schematically in Fig. 1c. The rotation matrix, $\mathbf{R}$, polarization vector, $\vec{E}$, and local orientation field, $\vec{o}$, are given in index notation, $R_{ij}$, $E_i$ and $o_j$, and Einstein summation notation is indicated with subscript indices for conciseness. Using the coordinate system we have just introduced, the X-ray polarization vector can be expressed as $\vec{E} = (\cos\alpha, \sin\alpha, 0)$, where $\alpha$ is the polarization angle, between the electric field direction and the $x$ axis, as indicated in Fig. 1c.

The projections, given by Equation 2, are defined in terms of the scattering factors, $f_0$ and $f_{\text{lin}}$. Depending on the imaging technique used, the measured data is related to these scattering factors in different ways. For example, with absorption-based techniques such as (S)TXM[30,36], the projections correspond to the transmitted intensity, $I(x, y)$. Alternatively, with coherent imaging techniques such as ptychography[37] and Fourier transform holography[38] both amplitude, $A(x, y)$, and phase, $\varphi(x, y)$, projections of the sample are recovered. Formally, these quantities can be linked to the linear dichroic projection, $P(x, y)$, defined in Equation 2 as follows:

$$I(x,y) = \exp(-2\pi\lambda r_e \Im\{P(x,y)\}),$$

$$A(x,y) = \exp(-\pi\lambda r_e \Im\{P(x,y)\}), \text{ and} \tag{3}$$

$$\varphi(x, y) = \exp(i\pi\lambda r_e \Re\{P(x, y)\}),$$

where $r_e$ is the classical electron radius, $\lambda$ is the wavelength of the incident X-rays, and $\Im$ and $\Re$ denote the imaginary (or dissipative) and real components of the complex projection, respectively.

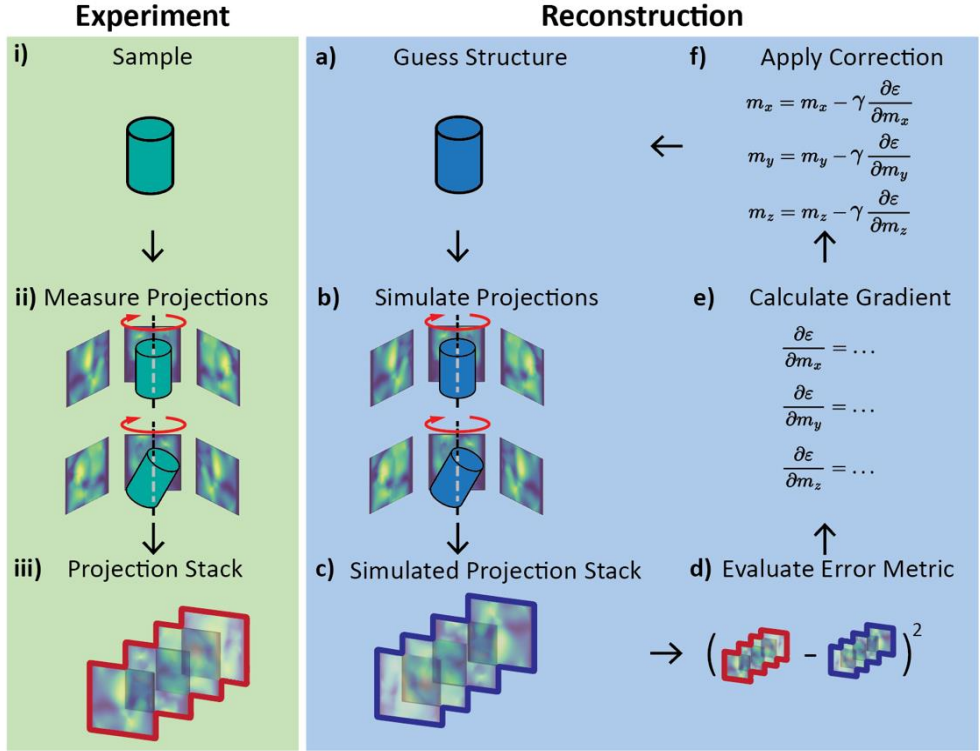

**Figure 2**: (i-iii) Experimental measurement of linear dichroic projections from the sample. (a-f) Protocol of the iterative algorithm for recovering the internal 3D configuration of the orientation in the sample.

## 2.3 Development and implementation of the XL-DOT reconstruction algorithm

In this section, we describe the construction of our iterative gradient-based reconstruction method that operates on the experimentally acquired linear dichroic projections. A schematic overview of the experimental measurement and the reconstruction algorithm is given in Fig. 2, with the experimental protocol illustrated in panels i-iii, and the reconstruction process detailed in panels a-f.

At the beginning of the reconstruction process, a random initial guess of the orientation, $\vec{o}$, within the sample in made (Fig. 2a). Sets of linear dichroic projections are then simulated from this guess for each axis about which the sample is rotated in the experiment, with two orientations of the sample (or tilt axes; $\phi = 0$ and $\phi > 0$) shown in Fig. 2b. Each image in the simulated projection stack (Fig. 2c) is compared to its corresponding image (i.e. obtained at the same sample tilt and rotation) in the experimentally obtained projection stack (Fig. 2iii). The square of their difference defines the error metric (Fig. 2d), which is used to quantitatively evaluate the similarity of the reconstructed projections to the experimental data. Subsequently, the gradient of the error metric with respect to the $x$, $y$ and $z$ components of the orientation field (Fig. 2e) is used to adjust the current guess (Fig. 2f), reducing the difference between the simulated and measured projections and by extension the error metric. After each iteration, involving stepping through panels a to f in Fig. 2, the reconstructed configuration becomes a better representation of the orientation field of the measured sample. This process is repeated iteratively until a predetermined number of iterations is executed, or a predefined value of the error metric or threshold in the gradient is reached.

To mathematically express the process outlined above, we first denote the error metric — which quantifies the similarity between the current guess and the experimental data — as the square of the differences between the simulated projections, $\hat{P}(x, y)$, and the experimentally obtained projections, $P(x, y)$:

$$\epsilon = \sum_{n,x,y} \left[ P^n(x, y) - \hat{P}^n(x, y) \right]^2. \tag{4}$$

Here, the index $n$ indicates the projection obtained at the $n^{\text{th}}$ sample orientation for a given sample tilt and rotation described by the $n^{\text{th}}$ rotation matrix, $\mathbf{R^n}$.

The error metric is then reduced using a gradient-based approach, updating the guess of the orientation to give a better agreement between the real and simulated projections, which in turn leads to a better agreement with the configuration of the orientation within the sample, provided a unique solution exists. Specifically, the analytical derivative of the error metric in Equation 4 is defined with respect to the parameters that will be reconstructed — i.e. the three ($x$, $y$ and $z$) components of the orientation field — and then used to apply a correction to the current guess of the internal orientation. The analytical expression for the gradient of the error metric with respect to the components of the orientation field can be explicitly written as:

$$\frac{\partial \epsilon(x, y, z)}{\partial o_k} = 4f_{\text{lin}}\left[\hat{P}^n(x', y') - P^n(x', y')\right]\left(R_{ij}^n o_j E_j\right)\left(R_{lk}^n E_l\right)$$

$$= 4f_{\text{lin}}\left[\hat{P}^n(x', y') - P^n(x', y')\right]\left(R_{1j}^n o_j \cos\alpha + R_{2j}^n o_j \sin\alpha\right)\left(R_{1k}^n \cos\alpha + R_{2k}^n \sin\alpha\right). \quad (5)$$

Here, the subscript index $k$ indicates the components $x$, $y$, $z$ of the current estimate of the orientation field, $\vec{o}$. The free parameters of the measurement — the angle of the polarization, $\alpha$, and the imaging geometry described by the rotation matrices $\mathbf{R^n}$ — are included in the gradient and directly determine the quality of the final reconstruction. As the orientation field $\vec{o}$, appears in Equation 5, it directly influences the evolution of the iterative process. This means that the initial guess must be non-zero. Otherwise, the gradient would evaluate to zero, preventing updates to the orientation field. The effect of the initial conditions on the resulting orientation tomogram is discussed in Section 3.3.

## 3. Simulations to optimize and determine limitations of XL-DOT

To assess the robustness of XL-DOT in reconstructing diverse orientation configurations, we simulate linear dichroic projections from test structures generated using numerical methods. These simulated projections are then used as an input to the XL-DOT reconstruction algorithm. The resulting reconstructions are then compared with the test structures, referred to as the ground truth or reference structures, allowing for an assessment of the accuracy and robustness of the reconstruction algorithm. Additional insights can be obtained by analyzing the spatial distribution of errors, which helps to identify potential inaccuracies or limitations of the reconstruction algorithm.

To define the type of test structure to use, we first consider the typical orientation configurations and textures found in physical systems, whose form depends strongly on the properties of the material system in question. For example, in polycrystalline samples, the crystallographic orientation tends to be uniformly orientated within a single crystallite, with each crystallite separated by grain boundaries. Uniform domains separated by boundaries are also found in ferroelectrics[6] as well as antiferromagnetic materials that exhibit strong magnetocrystalline anisotropy[14,39]. In contrast, in some systems there can be spatial variations in the orientation that are more gradual. This is, for example, the case for liquid crystals and polymer fibers[1], where the director tends to change continuously, except in the vicinity of topological defects, referred to as disclinations[40,41], where the orientation rapidly changes. Smooth variations in orientation can also be present in the Néel vector of antiferromagnets with low magnetocrystalline anisotropy, in which vortex and meron structures can be observed[13]. In the following discussion, both regimes — granular with high anisotropy and smoothly varying — are considered, and the ability of XL-DOT to recover the structure for each is evaluated.

To create reference model configurations for both types of systems, we employed the micromagnetic software MuMax3[42]. Specifically, we simulated two distinct orientation configurations: a system featuring smoothly varying orientations with multiple vortex-like defects, and a system with well-defined unidirectional domains separated by sharp domain walls, representing the granular configuration. The two configurations, which will be referred to as "smoothly varying" and "granular", are shown in Fig. 3a and Fig. 3f, respectively. A wedge has been removed from the structures to reveal the internal 3D structure.

### 3.1 Reconstruction under ideal measurement conditions

We begin by simulating XL-DOT under ideal measurement conditions i.e., utilizing the parameters that yield the best reconstruction (see Appendix A). Here, the dichroic projections were simulated for X-rays with both linear horizontal (LH, $\alpha = 0°$) and linear vertical (LV, $\alpha = 90°$) polarizations, and for triple-axis tomography, i.e. measurement comprising three sample tilts, namely $\phi = -30°$, $\phi = 0°$ and $\phi = 30°$. A total of 1080 projections were simulated, with 360 projections per tilt, comprising 180 projections with LH polarization and 180 projections with LV polarization (see first row of Table 1 for the imaging parameters). The configuration of the orientation within the model systems was then reconstructed in 3D using the iterative reconstruction algorithm described in the previous section. Specifically, 20 individual reconstructions with different initial random guesses were performed, with each reconstruction comprising 500 iterations, and the final reconstruction was obtained by taking the average of these reconstructions. We show later, in Section 3.3, how averaging over multiple reconstructions with random initial guesses improves the reconstruction quality.

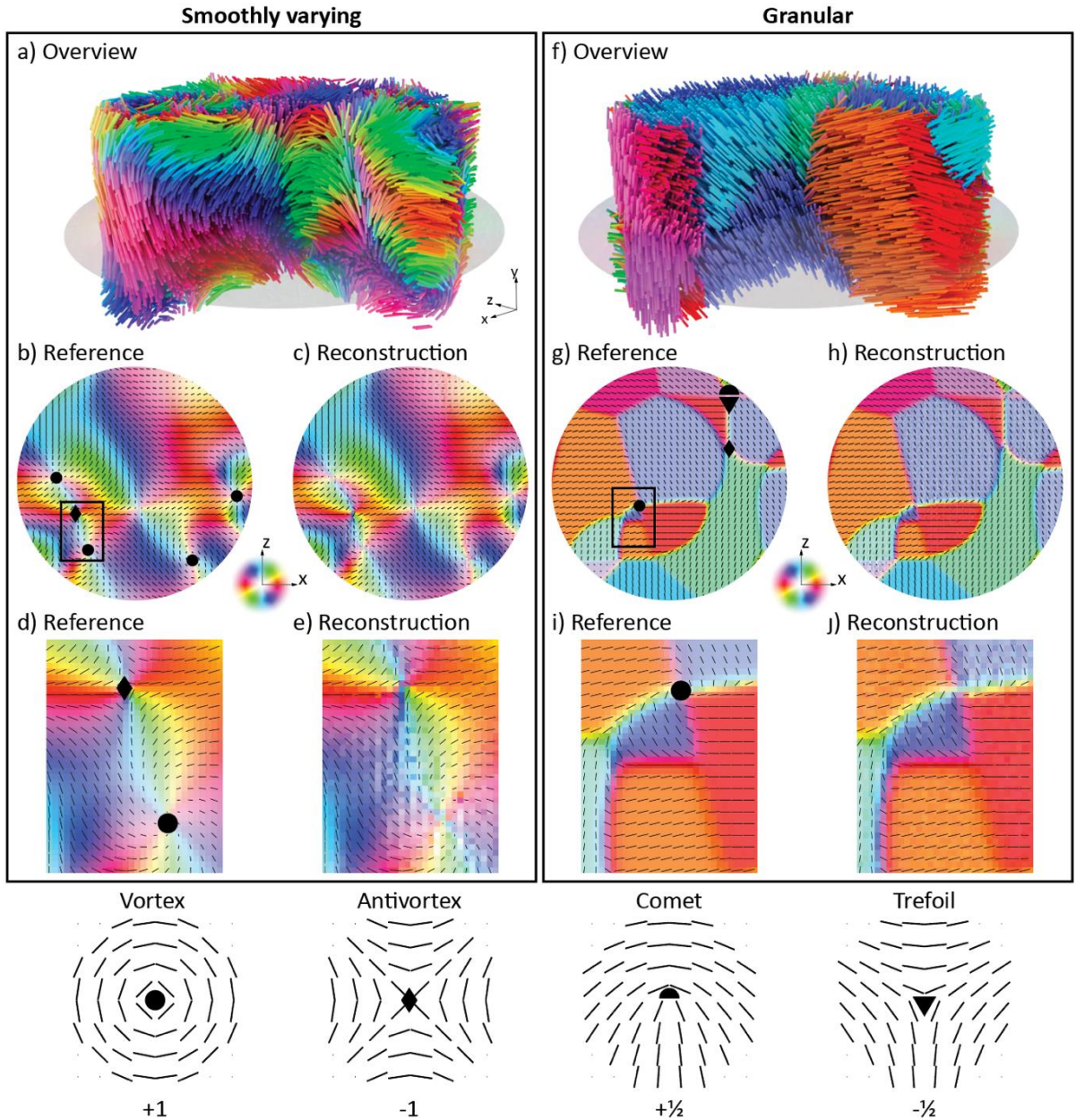

**Figure 3**: (a, f) 3D renderings of the a) smoothly varying and f) granular orientation configurations. (b, g) Reference and (c, h) reconstructed slices corresponding to the planes indicated in grey in (a) and (f). (d, e) Magnified view of the d) reference and e) reconstructed configurations hosting a vortex-antivortex pair taken from the smoothly varying configuration (region indicated with black rectangular box in panel b). (i, j) Magnified view of the i) reference and j) reconstructed configuration of the location where four domains meet taken from the granular configuration (region indicated with black rectangular box in panel g). Schematics topological features are shown below the main panels, with such features indicated in b, d, g, and i with black circles (vortices), diamonds (antivortices), semicircles (comets) and triangles (trefoils). The numbers under the schematics correspond to their winding number. Rods and their background color (see color wheel on coordinate axes) correspond to the orientation of the order parameter. Lighter hues signify an out-of-plane component (along the $y$ axis) in the orientation.

We determine the validity of the XL-DOT reconstructions in Fig. 3, by comparing them to the reference configurations. In general, the reconstruction qualitatively matches the reference structure in both the smoothly varying and granular regimes (see Fig. 3b, c and Fig. 3g, h). In particular, the orientation is correctly recovered in the vicinity of distinct features where the orientation changes rapidly, such as boundaries and also in regions of uniform alignment of the orientation.

For both regimes, topological textures can be observed. The recovery of such configurations poses a significant challenge for 3D reconstructions, as the orientation undergoes rapid variations around distinct features. These features may appear as points (e.g., diverging magnetic singularities called "Bloch points"[32]), lines (e.g., vortex tubes and disclinations), or surfaces (e.g., grain boundaries and domain walls). As a consequence, topological features represent challenging textures to recover both in terms of resolving them and obtaining the correct orientation. In the reference slices plotted in Fig. 3b, g, a number of topological features can be seen, which are categorized with different black shapes. Schematics of the topological features corresponding to these shapes with their winding numbers are shown in the lower part of Fig. 3. For example, on tracing a path around the core of a vortex (indicated using a black circle), the orientation rotates by 360° corresponding to a winding number of +1, while the orientation surrounding the antivortex core (indicated using a diamond) is saddle-like and has a winding number of –1. Accordingly, comet and trefoil defects have winding numbers of +1/2 and –1/2, respectively.

We first consider the smoothly varying regime, in particular the configuration shown in Fig. 3d, focusing on the region hosting a vortex-antivortex pair indicated by the black rectangle in Fig 3b. It can be seen that both the vortex and antivortex are correctly recovered in the reconstruction shown in Fig. 3d, maintaining their topological winding number and the overall orientation pattern, despite the presence of noise in the reconstruction.

We then consider the granular regime with the configuration of four domains that meet with a vortex topology, shown in Fig. 3i, which was taken from the region marked with a black rectangle in Fig. 3g. It can be seen that both the position and configuration of the domain walls is preserved in the reconstruction (Fig. 3j), as is the topology of the vortex structure.

Overall, this qualitative analysis gives a first indication of the ability of XL-DOT to successfully recover and map complex 3D orientation fields. We now proceed to a quantitative analysis by first defining an error parameter that gives a measure of the discrepancy between the reference configuration and the reconstructed orientation tomogram. Specifically, this metric quantifies the per-voxel angular difference between the reconstructed and reference structures, referred to as the "angular error" and defined as:

$$\epsilon_\theta = \arccos \frac{|\vec{o} \cdot \vec{o_r}|}{|\vec{o}||\vec{o_r}|}, \tag{6}$$

where $\epsilon_\theta$ is the per-voxel angular error, and $\vec{o}$ and $\vec{o_r}$ are the reference and reconstructed orientation fields, respectively.

Overall, we find that the reconstructions of the orientation configurations for both the smoothly varying and granular regimes exhibit a median angular error of 1.4° and 1.3°, and a 95[th] percentile angular error of 4.1° and 4.3°, respectively. These low errors for both regimes indicate a close agreement between the reference and reconstruction, demonstrating the overall success of the orientation reconstruction.

We then plot the spatial distribution of the angular error for the reconstructions in Fig. 4a-ii, d-ii where it can be seen that the angular error is highly localized to the vicinity of distinct features, such as the vortex-antivortex pair shown previously in Fig. 3e and the domain walls shown previously in Fig. 3j. In the vicinity of the vortex-antivortex pair (see Fig. 4a-iv), despite the fact that there are a handful of voxels (approximately 10) exhibiting an error greater than 30° and even up to 75°, the overall structure and topology of the configuration is preserved as the majority of voxels in the vicinity of these features have an angular error below 2°. A similar preservation of the structure is observed for the granular regime, where the angular errors are marginally small throughout the region of interest, as shown in Fig. 4d-iv, and the structure is correctly recovered.

## 3.2 Reconstruction under limited measurement conditions

Having ascertained that a successful reconstruction of complex orientation fields can be achieved for configurations in both granular and smoothly varying regimes, we proceed to investigate the impact of experimental constraints on the performance of the reconstruction method. This is because, at various beamlines or synchrotron facilities around the world, different combinations of X-ray polarization angles and tomographic tilts may be available for performing XL-DOT. We therefore examine the influence of the number of tomographic tilt axes and the number of X-ray polarization angles employed. In particular, we consider two cases: (i) dual-axis tomography, involving two orientations of the sample (as illustrated in Fig. 1c) with only LH and LV polarizations and (ii) single-axis tomography, where projections can be acquired at four intermediate polarization angles beyond the usual LH and LV. This is done as an attempt to mitigate the errors of using fewer tilts by including additional polarizations.

To allow for a direct comparison between different measurement configurations — including the triple-axis case described in Section 3.1, the dual-axis case, and the single-axis case — the total number of projections was kept constant. For dual-axis tomography, 1080 projections were split between two axes, with 270 projections per polarization (LH and LV) per tilt, and an angular spacing of 2/3°. For single-axis tomography, 1080 projections were split evenly between four polarization angles: 20°, 110°, 50° and 140° (see Appendix B), with 270 projections per polarization and 2/3° angular spacing between adjacent projections. The parameters used for triple, dual and single-axis tomography are summarized in Table 1.

| Label | Total projections | Tilt axes/ ° | Polarization angles / ° | Angular spacing / ° |
|---|---|---|---|---|
| Triple-Axis | 1080 (360 per tilt) | –30, 0, 30 | 0, 90 | 1 |
| Dual-Axis | 1080 (540 per tilt) | 0, 30 | 0, 90 | 2/3 |
| Single-Axis | 1080 | 0 | 20, 110, 50, 140 | 2/3 |

**Table 1**: Parameters used to simulate XL-DOT.

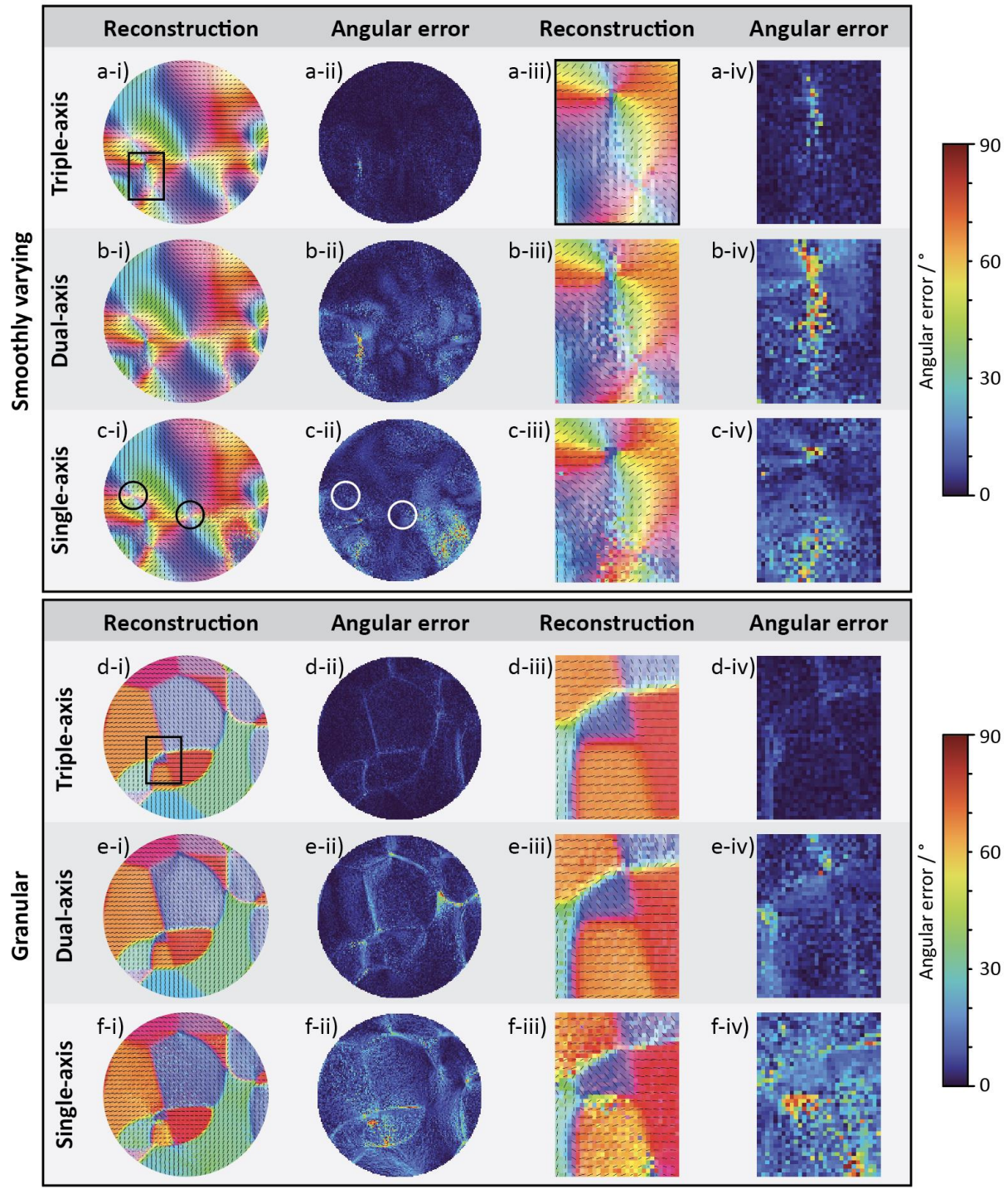

**Figure 4**: i) Orientation slices taken from the (a, d) triple, (b, e) dual and (c, f) single-axis XL-DOT reconstructions and their associated ii) angular error. iii) Magnified views of topological structures with the greatest angular error with respect to the reference structure, and iv) a map of the corresponding angular error.

We consider the quality of the reconstructions obtained from the triple, dual and single-axis geometries for the configuration in the smoothly varying regime in Fig. 4a, b, c, and for the configuration in the granular regime in Fig. 4d, e, f, respectively, where the reconstructed orientation and angular error are given for the whole slice (i, ii) and the magnified region (iii, iv). We note first that, for all three measurement geometries, the orientation is generally well reconstructed, with the overall configuration recovered in each case. However, when considering the reconstruction in more detail, one can see that, while the angular error for the triple-axis XL-DOT is negligible, reducing the number of tilt axes used in the tomographic measurement leads to the enhancement of the angular error, as can be seen by the increase in noise in the reconstructions (Fig. 4i,iii), and the increase in angular error in the vicinity of the defects with sharply changing orientation (Fig. 4ii,iv). The median angular error for the configurations in the smoothly varying and granular regimes is 3.6° and 2.7°, respectively for dual-axis tomography, which are larger than the 1.4° and 1.3° angular errors for the triple-axis reconstruction. For the single-axis tomography with four polarizations, the median angular error further increases for both the smoothly varying and granular structures to 6.7° and 6.0°, respectively. Reducing the number of tilt axes leads to a significant increase in the $95^{th}$ percentile error, i.e. the largest angular error that 95% of the voxels exhibit. In particular, the $95^{th}$ percentile error increases from approximately 4° for the triple-axis reconstruction to 13° for dual-axis reconstruction, and up to 24° for single-axis reconstruction. A low median angular error ($< 5°$) indicates that an accurate representation of the structure can be recovered, whereas a high $95^{th}$ percentile error ($> 15°$) suggests that there are regions present that are poorly recovered. For a straightforward comparison of the effect of the number of tilt axes on the distribution of angular errors in the entire structure, a summary is provided in Table 2.

Despite the increase in the error as the number of tilts is decreased, the low magnitude of these angular errors is representative of the fact that, overall, the configuration is well reconstructed even for fewer tilt axes. In particular, it is a promising sign that a representative reconstruction can be obtained, even with a single rotation axis, which potentially reduces the need for complex multi-tilt experimental setups. However, we note that, in order to obtain the highest quality reconstruction of the orientation field, at least three tilt axes would be required.

Qualitatively, angular errors in the reconstruction can be seen to be greater near regions of rapidly varying orientations, such as topological defects and domain walls, as depicted in Fig. 4ii. Upon closer inspection of the angular error in certain topological features in the configuration (Fig. 4iv), there appears to be a significant angular error of up to 90° near the core of a topological vortex in all reconstructions (Fig. 4a-c-iv), and in its immediate vicinity. Interestingly, other features are not impacted as significantly or at all. For instance, the hedgehog vortex structure (left black circle in Fig. 4 c-i) and an antivortex structure (right black circle in Fig. 4 c-i), have a low associated angular error (white circles in Fig 4c-ii).

| Structure | Method | Angular error / ° | |
|---|---|---|---|
| | | Median | 95th percentile |
| Smoothly varying | Triple-axis | 1.4 | 4.1 |
| | Dual-axis | 3.6 | 14.4 |
| | Single-axis | 6.7 | 26.5 |
| Granular | Triple-axis | 1.3 | 4.3 |
| | Dual-axis | 2.7 | 11.6 |
| | Single-axis | 6.0 | 22.4 |

**Table 2**: Reconstruction angular errors for the different reconstruction methods for configurations in both the smoothly varying and granular regimes.

This variation in reconstruction accuracy across different features suggests that the nature of the texture in the orientation field plays a critical role. In particular, it was previously observed for circular dichroic magnetic tomography that an enhancement of the error was associated with the presence of divergent configurations[33]. To test whether the nature of the orientation field — i.e. the local divergence and the local curl — affects the quality of the reconstruction, we simulated XL-DOT for a series of model vortex structures that exhibit different combinations of divergence and curl, shown schematically in Fig. 5b-e. According to their appearance, we refer to these structures as a conventional vortex (Fig. 5b), counterclockwise (CCW) helical vortex (Fig. 5c), hedgehog vortex (Fig. 5d) and clockwise (CW) helical vortex (Fig. 5e). Such planar configurations can be embedded in 3D configurations, for instance, as constituents of vortex tubes or at the center of Bloch points (see Fig. 5a). We note that all structures maintain the overall topology and winding number of a vortex but the balance of rotational (solenoidal, curl) and irrotational (divergence) field components are different. We now simulate XL-DOT for the geometry of single-axis tomographic imaging shown in Fig. 5a. We observe that, as these structures are rotationally symmetric, the same projection is obtained for all tomographic rotation angles $\theta$ when the rotation axis is orthogonal to the plane of the feature as seen in the sinograms in Fig. 5j-m. When attempting to reconstruct these structures (Fig. 5f-i), the limits of the reconstruction algorithm become evident: While the conventional and hedgehog vortices are well reconstructed, CCW and CW helical vortices cannot be reconstructed with single-axis tomography. This is because the same projections are obtained from the CCW and CW helical vortices (Fig. 5k, m), meaning that a unique solution does not exist, and the projections cannot be mapped onto a unique orientation field.

To determine whether this lack of a unique solution leads to the enhanced error observed for certain topological features in the numerical simulations, we consider the average error for voxels with different divergence and curl as shown in Fig. 5n. Specifically, the entire structure of the smoothly varying regime shown in Fig. 3a was grouped into bins according to the local divergence and magnitude of the curl, also referred to as the irrotational and solenoidal (rotational) components. We then plotted the average error exhibited by each group as a histogram in Fig. 5n. In the histogram, it can be seen that, although the error of purely irrotational or purely solenoidal voxels remains relatively low, voxels that have high magnitude of both the divergence and the curl exhibit a disproportionately large error. An example of the single-tilt reconstruction of a CW helical vortex, i.e. a feature that combines both high divergence and curl, is shown in Fig. 5o. Here it can be seen that the single-axis reconstruction is noisy, and can therefore conclude that, for single-axis orientation tomography, such in-plane helical textures with both curl and divergence represent a potential source of error. However, we note that as soon as an additional tilt axis is introduced, the local median angular error of such defects decreases, in this case from 21° to 6.2° (down to 2.4° when three tilt axes are used, as shown in Fig. 5p), indicating that, for experimental measurements, additional tilts will give sufficient information about the defects to correctly reconstruct them.

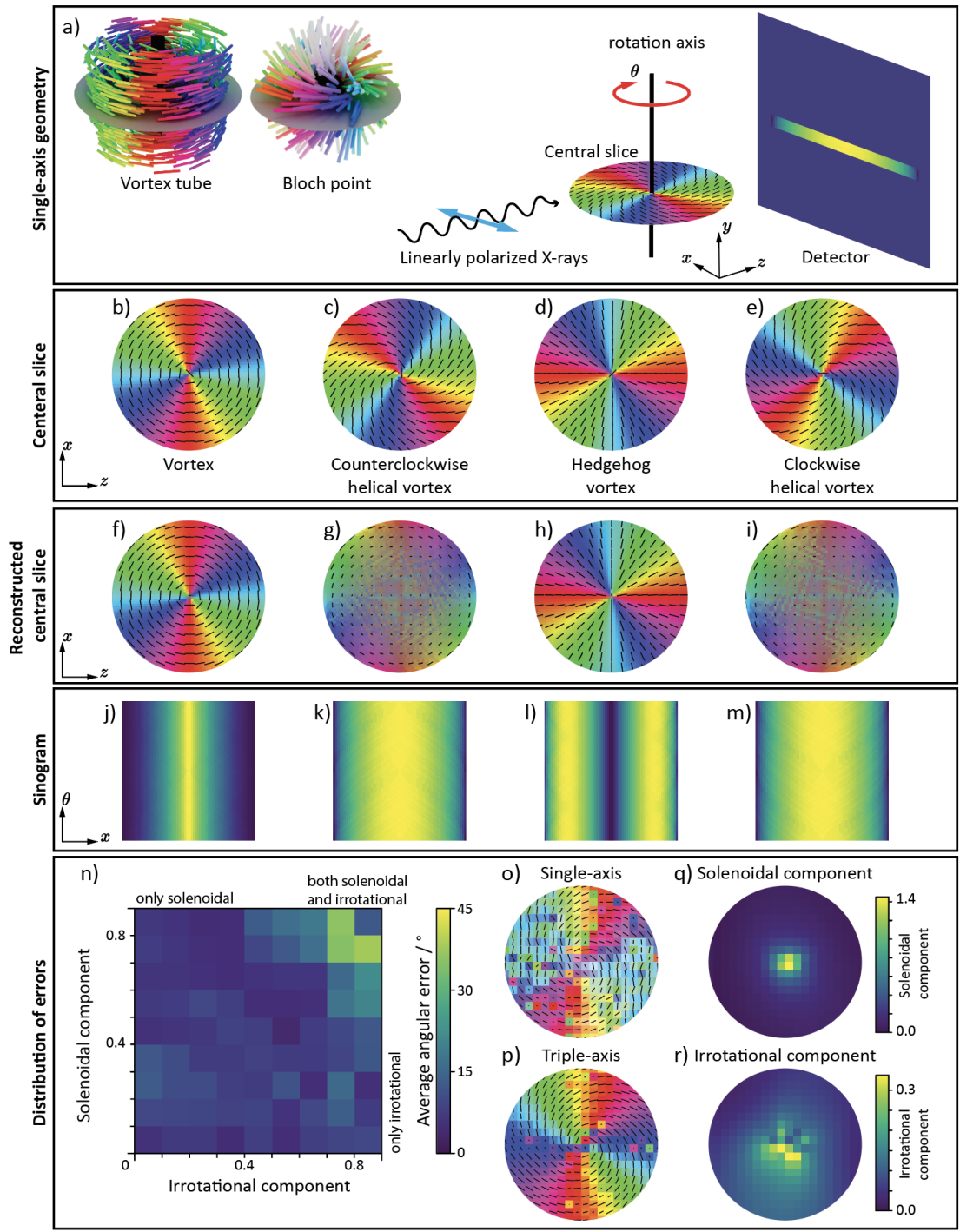

**Figure 5**: a) Geometry of single-axis XL-DOT, with the focus on the central slice of a Bloch point or vortex tube (top left insets) with vortex topology. (b-e) Different configurations with a winding number of +1: b) conventional vortex, c) counterclockwise (CCW) helical vortex, d) hedgehog vortex and e) clockwise (CW) helical vortex. (f-i) linear dichroic reconstruction for corresponding slices in b-e. While the f) conventional and h) hedgehog vortices are accurately reconstructed, the g) CW and i) CCW helical vortices are not recovered. To understand this result, the measured linear dichroic projections for each slice are plotted as a function of angle in (j-m) for the corresponding vortices, which are called sinograms. It can be seen that the sinograms from the CW and CCW helical vortices are identical. Thus, a unique solution is not attainable, and measurements with additional tilt axes are necessary to be able to distinguish between the two configurations in the reconstruction. n) Distribution of errors in the single-axis reconstruction of the smoothly varying structure shown in Fig. 3a. The average error is calculated for voxels exhibiting a given combination of irrotational (divergence) and solenoidal (rotational, curl) components, and a high error is seen in regions where both components are present. Examples of o) single-axis

and p) triple-axis reconstructions of a CW helical vortex obtained from the smoothly varying configuration, along with the corresponding magnitudes of the q) curl and r) divergence of the reference structure.

### 3.3 Effect of the initial guess on the reconstruction

Finally, we consider the impact of the initial guess on the outcome of the reconstruction process, as the initial guess of the orientation, $\vec{o}$, appears in the gradient (Equation 5), and therefore influences the reconstruction result.

One possibility is to choose the initial guess to have a uniform configuration, such that all voxels have the same starting orientation. Uniform initial configurations result in small errors in regions where the orientation varies smoothly, but can also cause the formation of fictitious structures (shown in Fig. C1b-d in the Appendix), which are likely to be a result of the convergence of the gradient descent algorithm to local minima. Such fictitious structures can be difficult to validate in experimental implementations of XL-DOT where the configuration in the sample is not known.

In order to avoid the formation of fictitious structures, we focus instead on using an initial guess where each voxel is assigned a random orientation. However, although the resulting reconstructions no longer suffer from the presence of fictitious structures, the random initial conditions lead to an increase in the angular error, even in smoothly varying regions (see Fig. C1e-g in the Appendix).

To reduce the reconstruction error and by extension improve the accuracy of the reconstruction, multiple tomograms, each obtained using unique, random initial conditions, can be averaged, as illustrated in Fig. 6a. Indeed, the averaging process reduces the angular error of the reconstruction, with the most pronounced improvement in both the median and 95[th] percentile angular errors observed when averaging five reconstructions. Overall, increasing the number of individual reconstructions being averaged leads to a further reduction in angular error, as indicated by the black trend line in Fig. 6b, albeit with diminishing returns.

This averaging approach also allows for the calculation of the standard deviation between the individually reconstructed tomograms, yielding a per-voxel estimate of the error in the reconstruction, which we refer to as the "uncertainty", highlighting regions of low confidence. Examples of slices of the uncertainty obtained through averaging the reconstructions are displayed in Fig. 6c-h, with the corresponding angular error in Fig. 6i-n.

The angular error is an important measure of the accuracy in the reconstruction. However, in the experiment it is not possible to calculate the angular error, but it is possible to look at the uncertainty. To assess the ability of the uncertainty to estimate the angular error, uncertainty and angular error maps were calculated for the reconstructed structures and are shown in Fig. 6c-h and Fig. 6i-n. Overall, the calculated uncertainty appears to overestimate the magnitude and presence of errors, sometimes highlighting regions that do not demonstrate significant angular error. Nevertheless, the uncertainty is moderately spatially correlated to the angular error, with Pearson correlation coefficients ranging from 0.46 to 0.60. Consequently, the uncertainty map can serve as a guide for experimental analysis by highlighting regions of low confidence that can be excluded from the analysis.

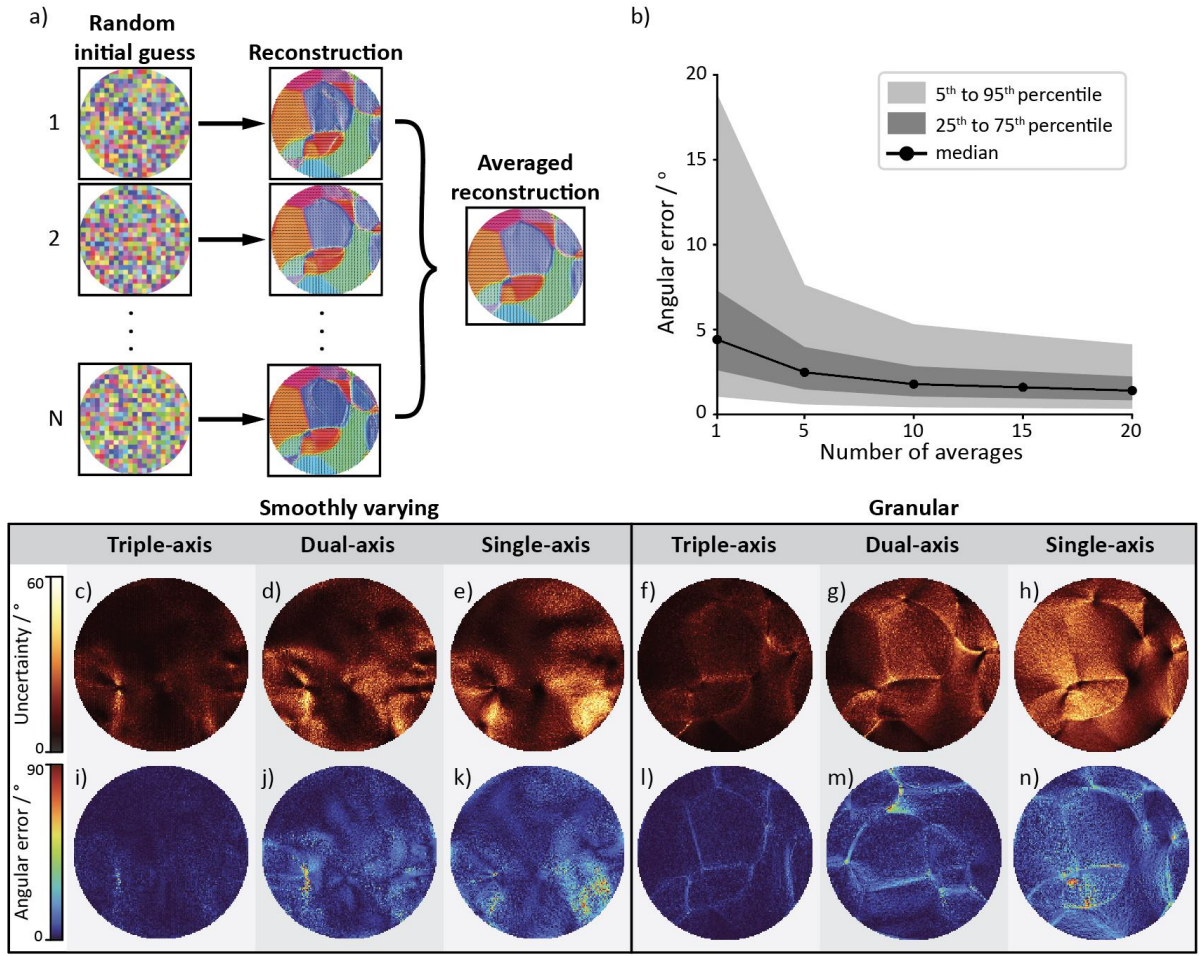

**Figure 6**: a) Illustration of the averaging procedure. For each reconstruction, a random initial guess is made with the orientation of each voxel chosen at random (voxel size exaggerated for clarity). b) Plot of angular error versus the number of averaged reconstructions demonstrating the effectiveness of the averaging process in reducing the angular error. The greatest improvement is on increasing the number of averages to five, after which the rate of improvement decreases. (c-h) Per-voxel uncertainty maps obtained from averaging 20 reconstructions. (i-n) Angular error maps corresponding to the uncertainty maps in (c-h), where a correlation between the uncertainty and true error is seen.

## 4. Conclusions and Outlook

In this work, we have presented the algorithm used for XL-DOT reconstruction and have shown that it can accurately recover the 3D orientation field of a probed magnetic or charge anisotropy in both smoothly varying and granular regimes. With numerical simulations, we have confirmed that the reconstruction process accurately recovers the probed orientation, with a median angular error of approximately 2° for a triple-axis measurement. This ideal case, measuring projections using three tomographic tilt axes and two polarizations (LH and LV), yields the most accurate reconstruction. When the number of tilt axes is limited to one or two axes, comparable results can still be achieved by including additional polarizations. However, this approach incurs increased angular errors that predominantly appear near textures such as grain boundaries and topological features that have a combination of curling and divergence in the orientation. These results indicate that even under non-ideal imaging conditions, XL-DOT can still give a consistent and accurate reconstruction of the orientation in 3D. Therefore, XL-DOT has potential applications in imaging a variety of orientation configurations in 3D, even with limited access to multi-tilt or variable polarization setups.

With the broad variety of applications — both in terms of different types of nanoscale configurations, as well as the broad variety of materials that exhibit linear dichroism — XL-DOT is a highly-promising tool for the future of material characterization. In particular, this technique opens the way to the imaging and understanding of hierarchical ordering in materials, which is essential for the development and improvement of multifunctional materials. Since linear dichroism is sensitive to the anisotropic distribution of charges or spin alignment in a broad spectrum of materials, XL-DOT can be used for the characterization of biological[11,12], structural[1,43,44], and quantum materials[13,14,45]. Finally, we note that XL-DOT can be combined with conventional transmission-based imaging techniques such as STXM[30,34] and coherent imaging techniques such as ptychography[26,37,46] and Fourier transform holography[38,47]. Indeed, coherent imaging techniques stand to benefit significantly from improvements in coherent flux with the next generation of synchrotrons. Here, the increased spatial resolution and the speed of data acquisition will make it possible to use XL-DOT to investigate nanoscale features across a wide range of materials, and track the orientation fields in situ over time on application of various stimuli or under different experimental conditions.

## Appendices

### Appendix A: Influence of number of tilt axes on the angular error

In vector tomography, rotating the sample with different tilt axes, with the tilt angle given by $\phi$ (see Fig. 1c) and recording sets of projections at different rotations $\theta$ is essential for obtaining all necessary information for a reconstruction. However, it is not clear how many tilt angles are necessary and which tilt angles yield the best results for XL-DOT.

To address this, XL-DOT was simulated including 720 projections, first for a single tomographic rotation axis — 360 with linear horizontal (LH) polarization and 360 with linear vertical (LV) polarization. This was repeated for up to 10 tilts per simulation, distributing the 720 projections evenly among the number of tilts in a simulation, with single-tilt tomography having 720 projections per tilt, dual-axis 360 per tilt, and so on. In this way, individual reconstructions were performed for the different numbers of tilt axes and the angular error was calculated in order to see how the number of tilts impacts the reconstruction quality. The results are shown in Fig. A1, with the 5th to 95th percentiles of the angular error colored in light grey, and the 25th to 75th percentiles colored in darker grey. The median angular error is depicted with a black line. From the graph, it can be understood that a single tilt with LH and LV polarizations leads to large errors in the reconstruction and, while at least two tilts lead to a significant decrease in the reconstruction error, three tilt axes result in optimal reconstruction quality with the lowest errors, with additional axes only marginally changing the quality of the reconstruction. For triple-axis tomography, the median angular error is low overall ($< 5°$), with at least 95% of the reconstructed angles matching those of the reference configuration to $4°$.

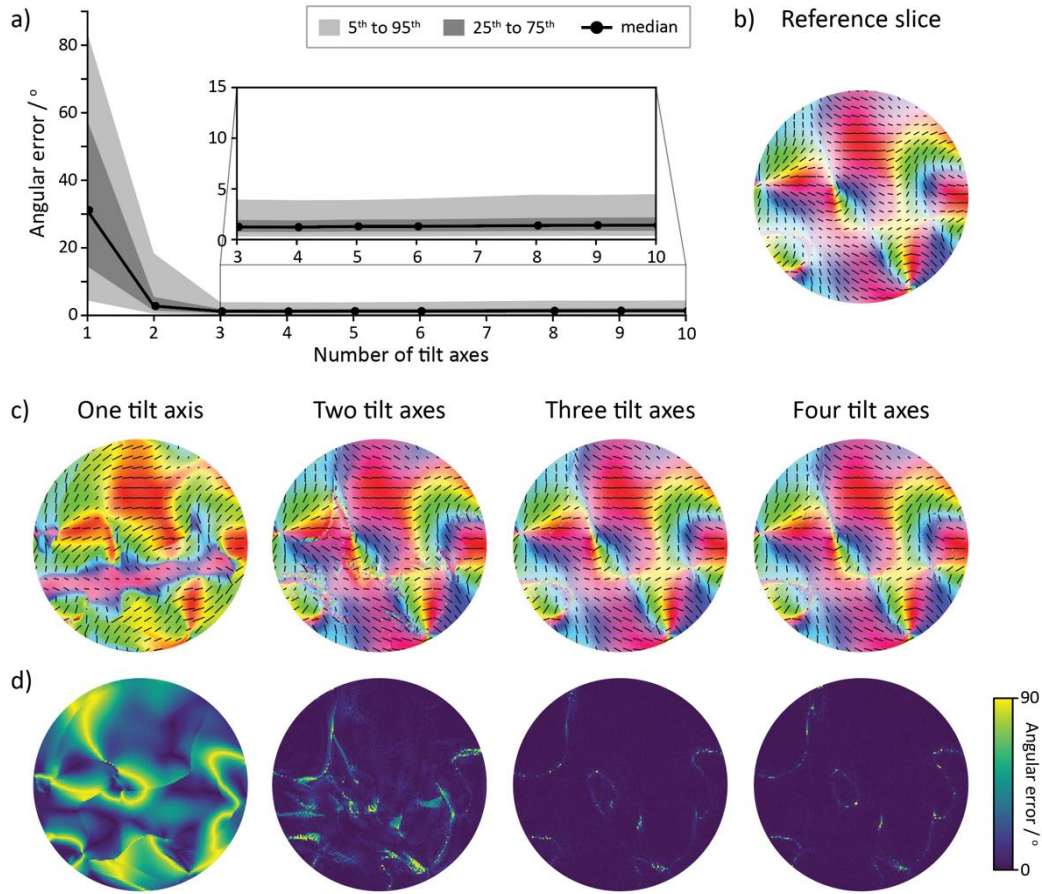

**Figure A1: Effect of number of tomographic tilt axes on the angular error.** a) Plot of angular error against number of tomographic tilt axes. The angular error calculated for each point corresponds to a single reconstruction with uniform initial conditions, i.e. all voxels were set to the same starting orientation. b) Example slice taken through the reference structure and c) corresponding slices obtained from reconstructions using one, two, three and four tomographic tilt axes. d) Per-pixel angular errors for the slices shown in (c).

## Appendix B: Single-axis tomography using four polarizations

In a similar way to changing the tilt angle, $\phi$, measurements at distinct polarization angles, $\alpha$, probe different combinations of the $x$, $y$ and $z$ components of the orientation and thus result in unique projections. This suggests that one might be able to obtain a reconstruction by merely changing the polarization angle of the beam, without the need for additional sample tilts. This can be beneficial for setups that are not adapted for multi-tilt tomography but where the polarization can be controlled using advanced insertion devices.

To identify the optimal parameters for this, tomographic reconstructions were performed using projections simulated for a single tilt axis, in particular $\phi = 0°$, but using two pairs of orthogonal polarizations. The results of the simulations are shown in the contour plot in Fig. B1. The plots are symmetric about the central diagonal (black dashed line in Fig. B1) as the two halves are equivalent. The two pairs of orthogonal polarizations that correspond to the lowest error, and therefore best reconstructions, are marked with crosses in Fig. B1 and occur for the polarization angle pairs of 20°, 110° and 50°, 140°. These were the polarizations used for the single-axis reconstruction described in Section 3.2. In the contour plots, a lower error is provided by four polarizations that are evenly spaced through 180° (so that the polarizations are 45° apart), and these are indicated with the white dashed lines in Fig. B1.

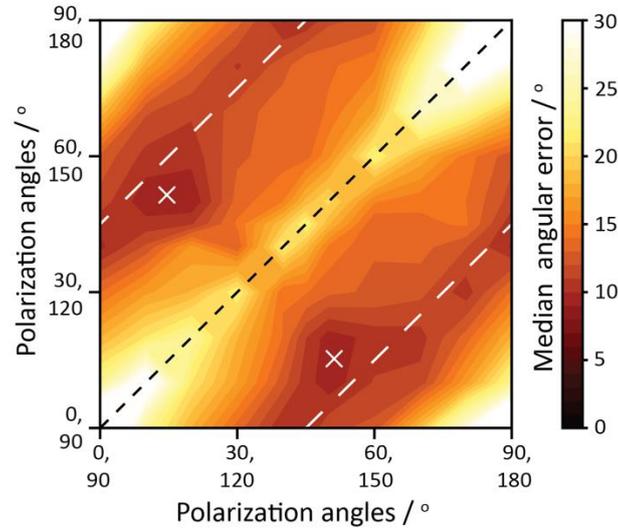

**Figure B1: Contour plot of median angular error for single-axis tomography with access to intermediate polarization angles (beyond LH and LV).** The projections simulated for two pairs of orthogonal polarizations (i.e. a total of four polarizations). The plot is symmetric about the dashed black line. Dashed white lines correspond to the polarization pairs that are 45° apart.

## Appendix C: Impact of initial conditions on the reconstructed structure

The initial guess for the orientation field has a direct impact on the resulting reconstruction. In the case of the dual-axis reconstruction, certain line artefacts appear to be present when using uniform initial conditions, shown in detail in Fig. C1b-d, and highlighted in the angular error map in Fig. C1b. In an experiment, it is not easy to validate these distinct structures as the configuration in the sample is not known. Such artefacts are likely to be formed when the gradient reconstruction relaxes to a local minimum, which is a common issue that occurs with gradient-based algorithms. By using random initial conditions instead, these fictitious structures disappear. However, this improvement is accompanied by an increased median error, jumping from 2.8° to 6.5°, indicated by the vertical red and blue lines in Fig. C1h, as well as an increased 95th percentile angular error from 18° to 37°, indicated by the vertical red and blue dashed lines in Fig. C1h.

To understand this difference in the error when selecting uniform as opposed to random initial conditions, we turn to the angular error distributions (or histograms), which are shown in Fig. C1h, i, with red and blue transparent bars corresponding to reconstructions performed using uniform and random initial conditions, respectively. In Fig. C1h, the errors distribution for the entire structure in the smoothly varying regime (Fig. 3a) is shown, with the error distribution exhibiting different widths and median values when using uniform and random initial conditions.

Since the imaging of topological structures and distinct features pose the most significant challenge to XL-DOT, a pair of histograms is also provided in Fig. C1i where only the voxels where the divergence in the orientation field is greater than 0.4 are taken into consideration. This provides more detailed information regarding the error near structures that are challenging both in terms of spatial resolution and accuracy of the reconstructed orientation. For this second pair of histograms, the median angular error is higher for both uniform and random initial conditions relative to the entire structure median values (compare vertical red and blue vertical lines in Fig. C1i with those in Fig. C1h). This rightward shift of the median angular errors indicates that the angular errors are not evenly distributed across the structure but are instead concentrated in regions where the orientation changes rapidly, such as near topological features. In addition, the histograms become more similar in terms of both their width and peak position, implying that (i) the two different initial conditions perform similarly in the vicinity of rapid changes in the orientation and (ii) the lower angular error when using uniform initial conditions can largely be attributed to regions where the orientation is smoothly varying. This is further highlighted by the absence of significant errors within the white dotted circle in Fig. C1b, which corresponds to a region devoid of distinct features. In contrast, these regions exhibit higher angular errors when random initial conditions are used (see region within white dotted circle in Fig. C1e).

As previously mentioned, using random initial conditions leads to a larger median angular error, but the reconstruction is less prone to producing fictitious objects such as the one shown in Fig. C1d, whose validity can be difficult to discern when reconstructing experimental data. Another advantage of using random initial conditions is that multiple reconstructions with different random initial conditions can be averaged, which both lowers the angular error as local minima are statistically eliminated, and provides sufficient per-voxel statistics to identify regions where the reconstruction is otherwise uncertain, as discussed in Section 3.3 of the main text.

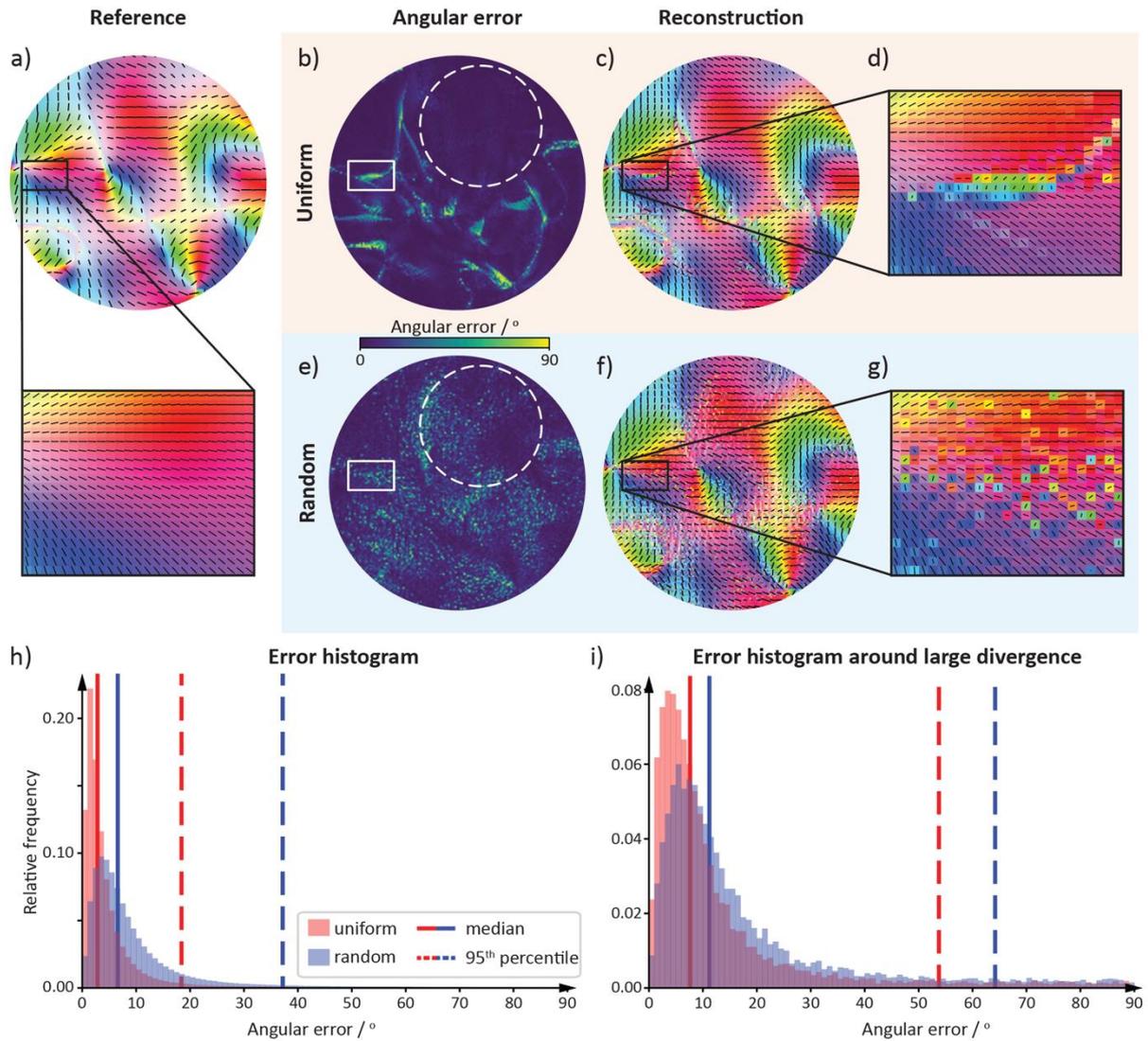

**Figure C1:** a) Slice from reference structure. Single reconstruction using (b-d) uniform initial conditions (highlighted with red background) and (e-g) random initial conditions (highlighted with blue background), both with two tilt axes. b, e) Per-voxel angular error for the corresponding (c, f) reconstructed slice with respect to the reference structure shown in (a). d) Isolated fictitious structure present in the reconstruction with uniform initial conditions and g) the same region obtained using random initial conditions provided for comparison. h) Angular error histograms for entire structure. i) Angular error histograms for voxels with divergence greater than 0.4. The continuous and dashed vertical lines indicate the median and 95[th] percentile errors, respectively. The red and blue colors indicate the use of uniform and random initial conditions, respectively.

## Acknowledgements

The research leading to these results has received funding from the Swiss National Science Foundation (SNSF), with project number 200021_192162, the Max Planck Society Lise Meitner Excellence Program and the European Research Council (ERC) under the ERC Starting Grant No. 3DNANOQUANT 101116043.